\documentclass[aps,prb,twocolumn,showpacs,superscriptaddress]{revtex4-2}  

\usepackage{bm}        
\usepackage{amssymb}   
\usepackage{color}
\usepackage{units}
\usepackage{amsmath}
\usepackage{natbib}
\usepackage{esint}
\usepackage{ulem}
\usepackage[english]{babel}
\usepackage[colorlinks=true, citecolor=blue]{hyperref}
\usepackage{xcolor}
\usepackage{chemformula} 
\usepackage[T1]{fontenc} 

\begin{document}
	\title{Hidden electronic phase in strained few-layer 1T-\ch{TaS2} }
	\author{Sruthi S}
	\affiliation{Department of Physics, Indian Institute of Science, Bangalore, Karnataka, India 560012}
	\author{Hemanta Kumar Kundu}
	\affiliation{Department of Physics, Indian Institute of Science, Bangalore, Karnataka, India 560012}
	\affiliation{Weizmann Institute of Science, Rehovot 7610001, Israel}
	\author{Prasad Vishnubhotla}
	\affiliation{Department of Physics, Indian Institute of Science, Bangalore, Karnataka, India 560012}
	\author{Aveek Bid}
	\email[E-mail:]{aveek@iisc.ac.in}
	\affiliation{Department of Physics, Indian Institute of Science, Bangalore, Karnataka, India 560012}

	\begin{abstract}
		
		Layered van der Waals materials are exciting as they often host multiple, competing electronic phases. This article reports experimental observation of the co-existence of insulating and metallic phases deep within the commensurate charge density wave phase in high-quality  devices of few-layer 1T-\ch{TaS2}. Through detailed  conductance fluctuation spectroscopy of the electronic ground state, we establish that the mixed phase consists of insulating regions surrounded by one-dimensional metallic domain walls. We show that the electronic ground state of 1T-\ch{TaS2} can be affected drastically by strain, eventually leading to the collapse of the Mott gap in the commensurate charge density wave phase. Our study resolves an outstanding question, namely the effect of the inter-layer coupling strength on the electronic phases in layered van der Waals materials.
		
	\end{abstract}
	
	\maketitle
	\section{Introduction}
	
	Transition metal dichalcogenides (TMD) are layered materials that have drawn the research community's attention due to their unique electrical, mechanical, and optical properties. Among this class of materials, 1T-\ch{TaS2} has been the subject of intense studies due to its rich electronic phase diagram. Multiple Charge density wave (CDW) transitions are present in 1T-\ch{TaS2} over a wide temperature range. Upon cooling, the system first undergoes a normal state to incommensurate CDW (IC-CDW) phase. With a further reduction in temperature, the system goes from IC-CDW to a nearly commensurate CDW (NC-CDW) phase and then on to an insulating commensurate CDW (C-CDW) phase~\cite{thomson1994scanning,wilson1975charge,rossnagel2011origin}. In the C-CDW phase, 13 Ta atoms come together to form a star of David arranged in a ($\sqrt{13}$ x $\sqrt{13}$)  superstructure. 
	
	{The thickness and cooling rate play a vital role in the NC-CDW to C-CDW transition~\cite{yoshida2014controlling}. As the thickness reduces, the NC-CDW to C-CDW transition is suppressed, and a supercooled phase of NC-CDW gets stabilized at low temperatures. A detailed study of the thickness and rate dependence was done by Yoshida et al. ¬\cite{yoshida2015memristive}. They show that the ordering kinetics of the phase transition can be tuned by changing the thickness. The reduction in thickness results in slow ordering kinetics, which results in the emergence of metastable states.}	
	
	Recently, the possible origin and electronic properties of the low-temperature insulating state in  1T-\ch{TaS2} has come under intense scrutiny. Initially, it was thought to be a Mott insulator~\cite{fazekas1979electrical,perfetti2005unexpected,kim1994observation,sato2014conduction} -- this was based on the presumption that inter-layer interactions could be ignored in comparison to the intra-layer interaction terms. Based on this, 1T-\ch{TaS2} was considered as a prime candidate to host quantum spin liquid state~\cite{law20171t}. However, recent theories that take into account inter-layer interactions propose that the ground state of 1T-\ch{TaS2} is instead a band insulator. Two exceptionally stable CDW configurations were identified in this state using DFT calculations. The ground state has a double layer stacking configuration where the dimerization of adjacent \ch{TaS2} layers make the system insulating. However, there is a metallic phase quite close in energy to this insulating phase. The energy difference between these two phases was estimated to be relatively small, leading to the suggestion that the system can be switched from one phase to another using external perturbations like pressure or doping.~\cite{lee2019origin,ritschel2018stacking,ritschel2015orbital}.
	
	Despite these advances, there are several open questions regarding the interplay of this system's various competing low-energy phases. The low-temperature C-CDW phase can be suppressed using various external perturbations~\cite{lahoud2014emergence,sipos2008mott,di1975effects,ang2012real,stojchevska2014ultrafast,vaskivskyi2015controlling}. Experiments, especially scanning tunneling microscopy measurements, show that the nature of the insulating ground state is strongly dependent on the inter-layer interactions and stacking order in 1T-\ch{TaS2}~\cite{butler2020mottness}. This implies that the low-temperature electronic phase of 1T-\ch{TaS2} can be modified by various parameters such as intercalation, chemical doping, and pressure leading to the appearance of a hidden state or metallic states~\cite{lahoud2014emergence,stahl2020collapse}. Recent STM measurements suggest strain as a possible tuning parameter that can induce a collapse of the Mott gap~\cite{bu2019possible}. To date, there are no clear signatures of the presence of these multiple phases from electrical transport studies. 
	
	In this article, we report the results of our study of resistance fluctuations in few-layer laterally strained 1T-\ch{TaS2} devices. We find that in strained flakes, the system goes into a metallic state below the NC-CDW to C-CDW transition temperature. Further, on heating, the system goes from a metallic state to an insulating state. We used low-frequency resistance fluctuation spectroscopy~\cite{ghosh2004set} to probe the underlying charge transport in the system. We find clear evidence that over a broad range of temperatures, the ground state of strained 1T-\ch{TaS2} consists of two distinct electronic phases. 
	
	\section{Results and Discussions}
	
\begin{figure}[t]
		\begin{center}
			\includegraphics[width=\columnwidth]{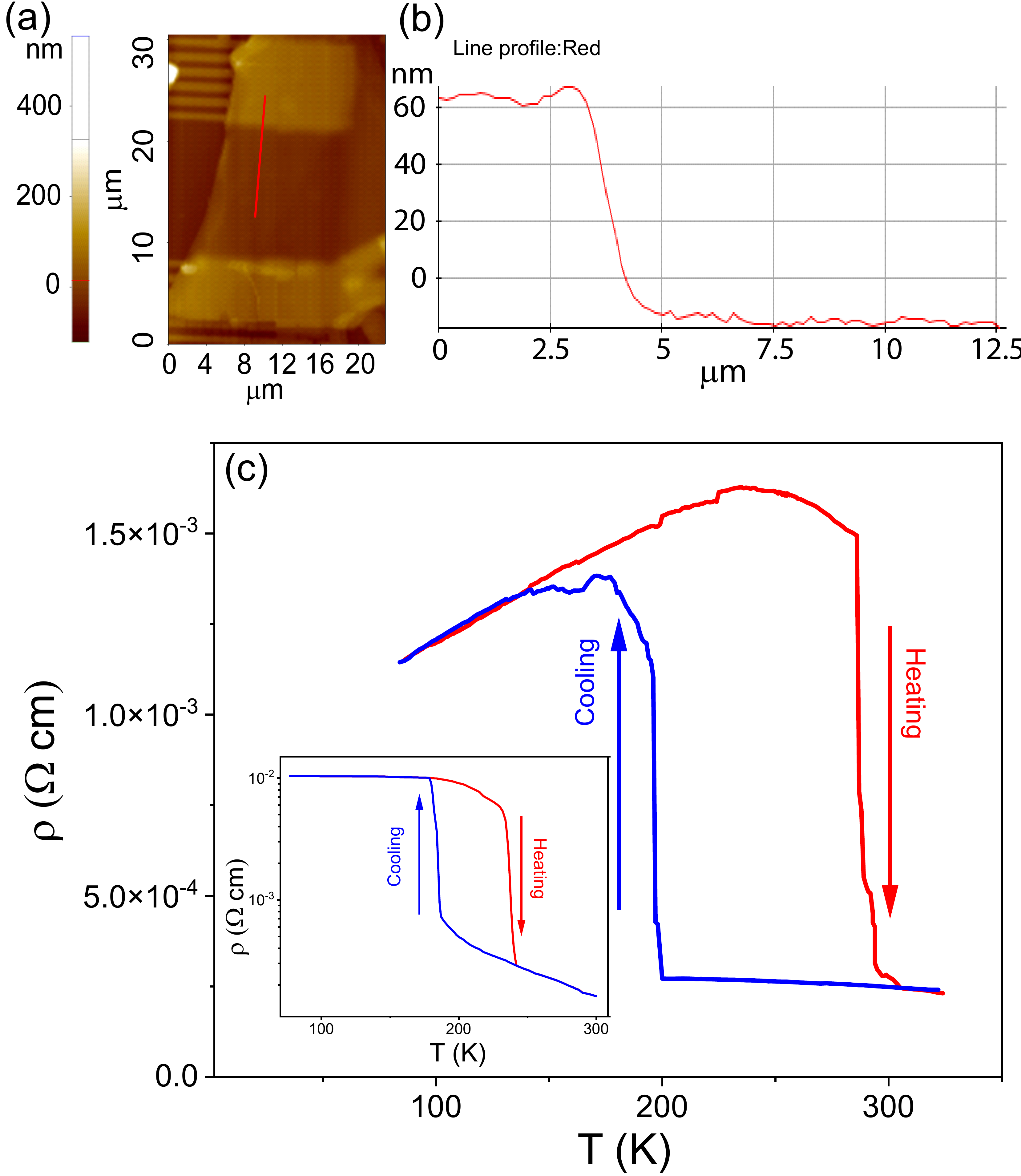}
			\caption {(a) AFM topography image of the device.(b) A topographic line profile is shown measured along the red
				line shown Fig~\ref{Fig.1}(a).(c) Plot of the resistance of the suspended  device versus measured Temperature during cooling (blue line) and heating (red line) runs. The inset on the left bottom shows a schematic diagram of  the resistance of the substrated device versus temperature~\cite{yoshida2014controlling,yoshida2015memristive,tsen2015structure}   \label{Fig.1}}   
		\end{center}
	\end{figure}

	1T-\ch{TaS2} flakes were mechanically exfoliated using scotch tape on silicone elastomer polydimethylsiloxane (PDMS) inside a \ch{N2} filled glove box with \ch{O2} concentration \textless0.1ppm. The thickness of the flakes was identified through optical contrast. Electrical contacts were prefabricated on Si$^{++}$/SiO$_2$ substrate using e-beam lithography followed by thermal deposition of Cr (5nm) and Au (60nm) films. Using an optical microscope-based transfer setup, the exfoliated flake was transferred from the PDMS on the pre-patterned Au probes. This process results in the 1T-\ch{TaS2} flakes being suspended on gold probes. {At low temperatures, there is a mismatch between the contractions of the Au pads and the 1T-\ch{TaS2} flakes  -- this results in a lateral strain in this device of about 1\%~\cite{kundu2017quantum,PhysRevB.82.155432}. Atomic Force Microscopy (AFM) topography of the flake (Fig.~\ref{Fig.1}(a)) shows it to be uniform with a thickness of  $\sim25$~nm.} 
	
\begin{figure}[t]
		\begin{center}
			\includegraphics[width=\columnwidth]{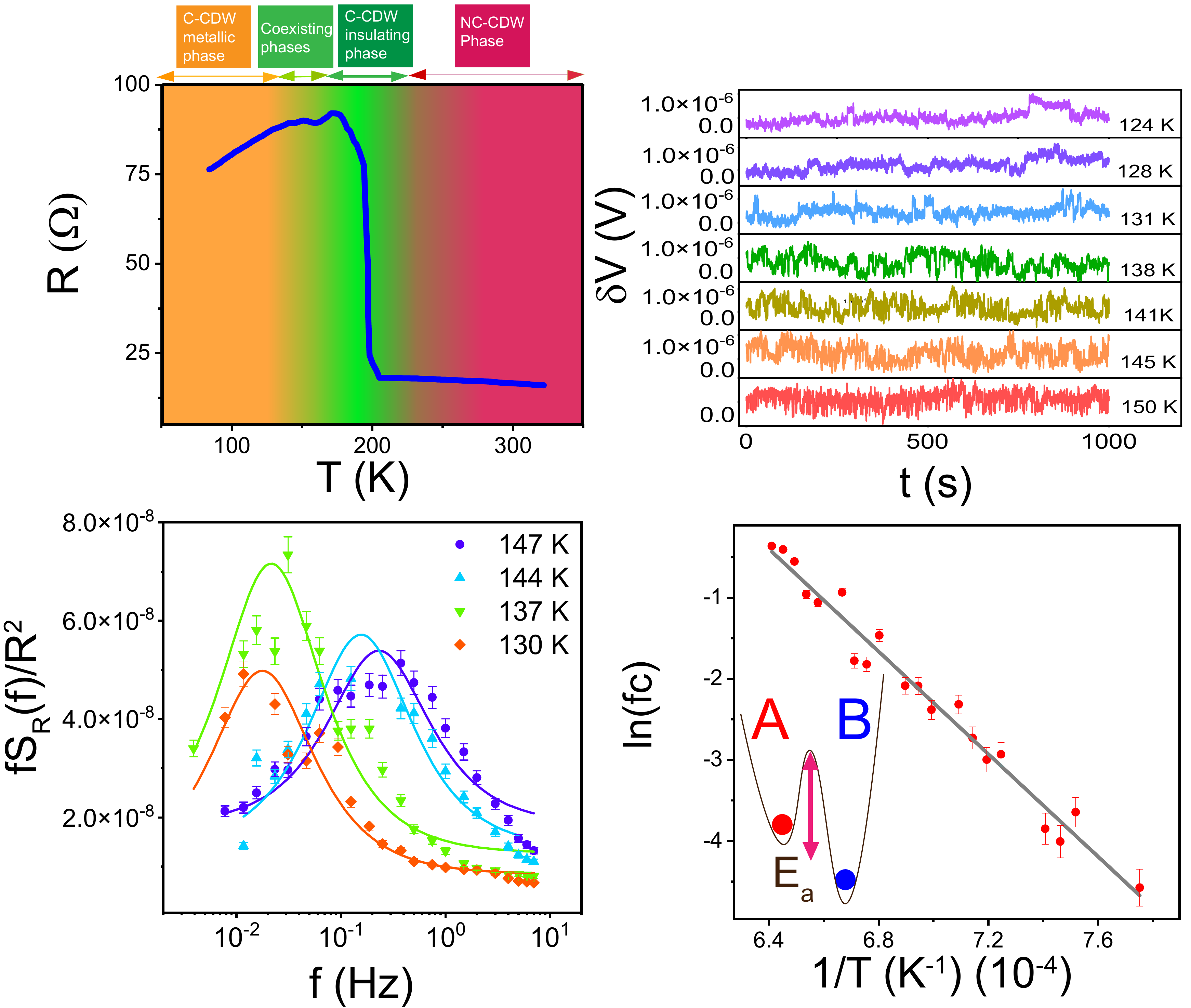}
			\caption{(a) Plot of resistance versus temperature while cooling which shows different phases in the system.  (b) Plot of the time-series of voltage fluctuations measured during the  cooling run.  (c) The scatter plots show the plots of the scaled PSD $S_R(f)/R^2$ as function of $f$ at the temperatures corresponding to the time series in (a). The solid lines corresponds to the fits to Eq.~\ref{eqn.1}.(d) Plot of $f_c$ as a function of $1/T$  -- the solid line shows the fit corresponding to Arrhenious relation Eqn.~\ref{eqn.2}.Inset shows a schematic representation of two phases  represented by A and B which are separated by an energy barrier $E_{a}$.  \label{Fig.2}}   
		\end{center}
	\end{figure}

	The temperature dependence of the resistance of the device over the temperature range 322~K$<T<$86~K is shown in the main panel of Fig.~\ref{Fig.1}(c). While cooling,  the large jump in resistance at $T_{max}^c=200$~K indicates a  phase transition from the nearly commensurate charge density wave to commensurate charge density wave. While heating, the reverse transition occurs around $T_{max}^h=286$~K. This large hysteresis in the resistive transition around this temperature range has been observed before, although the difference between $T_{max}^c$ and $T_{max}^h$ is unusually large in our case. {Recall that the transition between the NC-CDW and the C-CDW phases is a first-order phase transition, and as such, the transition between two phases shows hysteresis upon heating and cooling because of the presence of metastable states~\cite{yoshida2015memristive,tsen2015structure}.} {We checked the resistance versus temperature curve at different cooling rates, and we find that the hysteresis window remains same within $\pm$10~K with very little difference in the transition temperature.}  The resistance temperature plot shows that in contrast to on-substrate 1T-\ch{TaS2} flakes (shown in the inset of Fig.~\ref{Fig.1}(c))~\cite{yoshida2014controlling,yoshida2015memristive,tsen2015structure}, the suspended 1T-\ch{TaS2} flake remains in a metallic state (with positive $dR/dT$) down to at least 85~K, establishing that the insulating phase does not set in till very low temperatures in the suspended device. { Fig.~\ref{Fig.2}(a) shows the different phases in the suspended device of 1T-\ch{TaS2} while cooling.}  
	
	To probe the statistics of charge transport in the strained few-layer 1T-\ch{TaS2} devices, we performed low-frequency resistance fluctuation spectroscopy over the temperature range 86--324~K using a four-probe digital signal processing technique. A dual-channel lock-in-amplifier (LIA) was used to bias the device with a current of 100~nA at a carrier frequency 228~Hz. A low noise preamplifier (SR552) was used to amplify the voltage drop across the device, which the LIA subsequently detected. {To minimize the exposure of the device to the ambient, it was loaded in a cryostat within 30 minutes of the final fabrication step, and the cryostat pumped down to a pressure of $10^{-5}$~mbar.  The cryostat was cooled down in a liquid nitrogen Dewar. The temperature of the cryostat insert was controlled to better than 1 ~mK using a temperature controller.}  At several temperatures, during both the heating and the cooling cycles, the voltage data was recorded using a 16-bit analog-to-digital convertor card at a sampling rate of 2048 points/s over 32 minutes. During this time interval, the temperature was stable with $\delta T < 1$~mK.  The time series of voltage fluctuations $\delta v(t)$ was then digitally filtered and decimated to remove artifacts arising from aliasing and finite measurement bandwidth. The filtered time-series was then used to calculate the power spectral density (PSD) of the resistance fluctuations, $S_R(f)$ using the method of Welch periodogram over the frequency window 4~mHz<$f$<7~Hz. Thermal noise measurements of a standard resistor were used to calibrate the setup before actual measurements on the suspended 1T-\ch{TaS2} device. {The measurements were repeated for multiple cooling cycles – the data in each case was quantitatively similar. Our device, being a mesoscopic system, each thermal cycling reconfigures the conduction channel's defect configuration, making such measurements equivalent to ensemble averaging for all practical purposes.}
	
	Fig.~\ref{Fig.2}(b) shows plots of the voltage fluctuations $\delta V$, recorded during the cooling run of the device, at a few representative temperatures over the range 150~K to 122~K. {The resistance fluctuations was calculated by using the relation $	\left<\delta R^2\right> =  	\left<\delta V^2\right>/\left<I^2\right>$, where $\left<I\right> = 100$~nm is the source-drain current.} We find that the resistance fluctuates between two levels. The presence of this random telegraphic noise (RTN) indicates that the system is electronically phase-separated  -- it has access to two, nearly energetically equivalent phases that are separated by an energy barrier (see the inset of  Fig.~\ref{Fig.2}(d) for a schematic). {Note that these phases are not due to NC-CDW to C-CDW transition since that
		 transition has already occurred at 200~K.}
	
	In Fig~\ref{Fig.2}(c), we plot the power spectral density over the temperature range where RTN is present. The data have been multiplied by the frequency $f$ for the following reason: if the PSD  has a $1/f$ dependence, the quantity $fS_R\left(f\right)/<R>^{2}$ will be independent of frequency. On the other hand, the PSD of a time-series with random telegraphic noise is a Lorentzian with a characteristic frequency $f_c=1/\tau_c$, where $\tau_c$  is the timescale for switching between two levels. Motivated by this, we fit the PSD of the resistance fluctuations with the equation:		
	\begin{equation}
		S_R \left(f\right)/R^2 = {A/f+(Bf_c)/(f^2+f_c^2 )} 
		\label{eqn.1}   
	\end{equation}
	where $A/f$ is the $1/f$ component and $Bf_c/(f^2+f_c^2 )$ is the Lorentzian component of resistance fluctuations. $A$ and $B$ are constants and are determined from the fits to the experimental data. {The appearance of RTN in the time-series of resistance fluctuations and a corresponding Lorentzian component in the power spectral density indicates that there are two metastable states present in the system which are close by in energy (see the inset of Fig.~\ref{Fig.2}(d)). Thermal fluctuation-induced switching of the system between these two states leads to the appearance of the RTN.  As the process is thermally activated across an energy barrier $E_{a}$, the rate of transition between these two states is described by the Arrhenius relation.} We find $f_c$ to have a thermally activated behaviour of the form~\cite{kundu2017quantum,bid2003low,amin2015effect}:  
	\begin{equation}
		f_c={f_0 e^{-E_a/K_BT}} 
		\label{eqn.2}
	\end{equation}
	Fig.~\ref{Fig.2}(d) shows the plot of $ln f_c$ vs $1/T$. The activation energy derived from the slope of this plot was $272\pm 5$~meV. This energy scale is similar to the C-CDW gap shown through various experiments~\cite{lutsyk2018electronic,ang2012real}. This indicates that one of the phases is the C-CDW phase.

	\begin{figure}[t]
		\begin{center}
			\includegraphics[width=0.75\columnwidth]{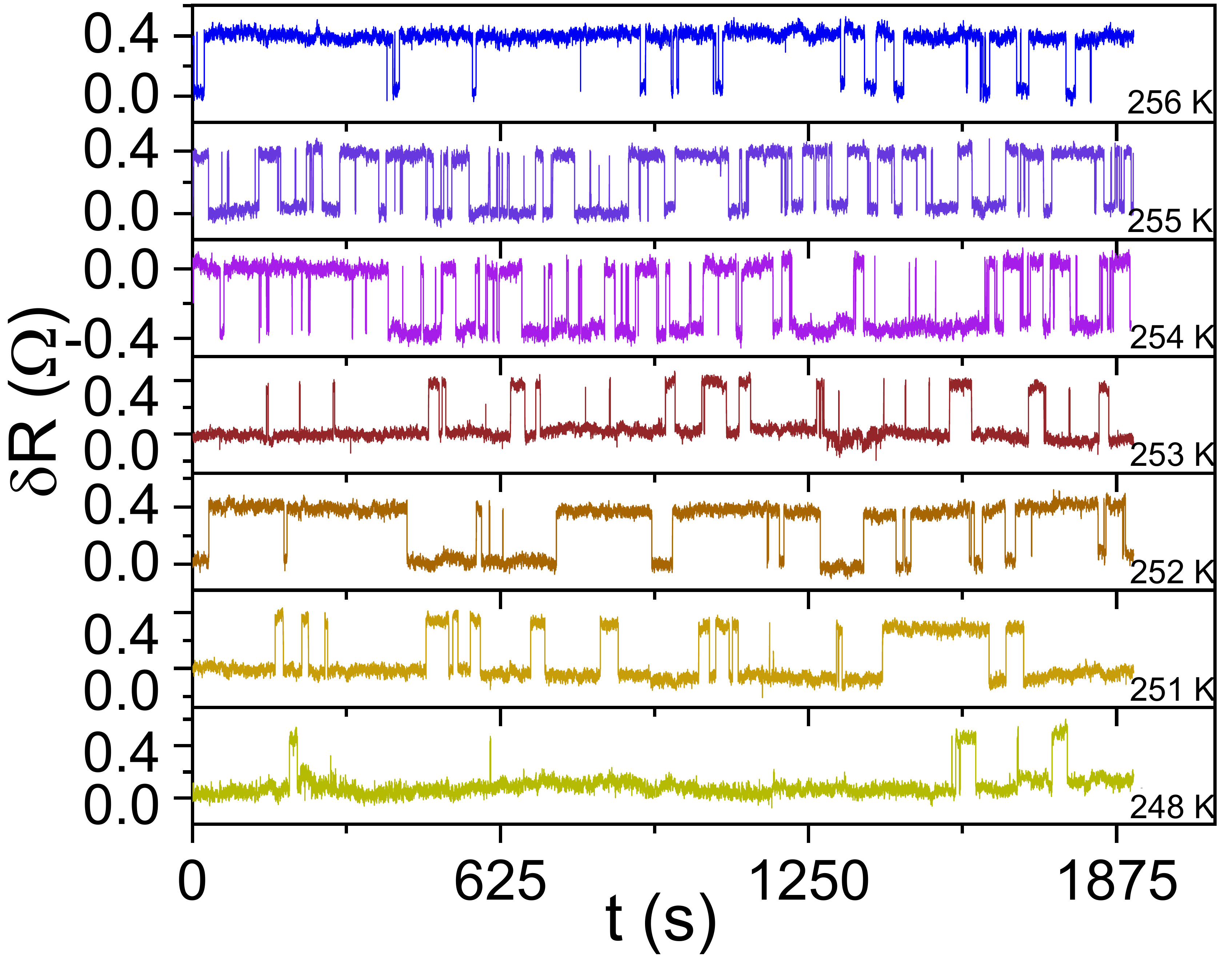}			
			\caption{(a) Plot of the time-series of resistance fluctuations measured during the  heating run.  
				\label{Fig.3}}   
		\end{center}
	\end{figure}

\begin{figure}[t]
	\begin{center}
		\includegraphics[width=1\columnwidth]{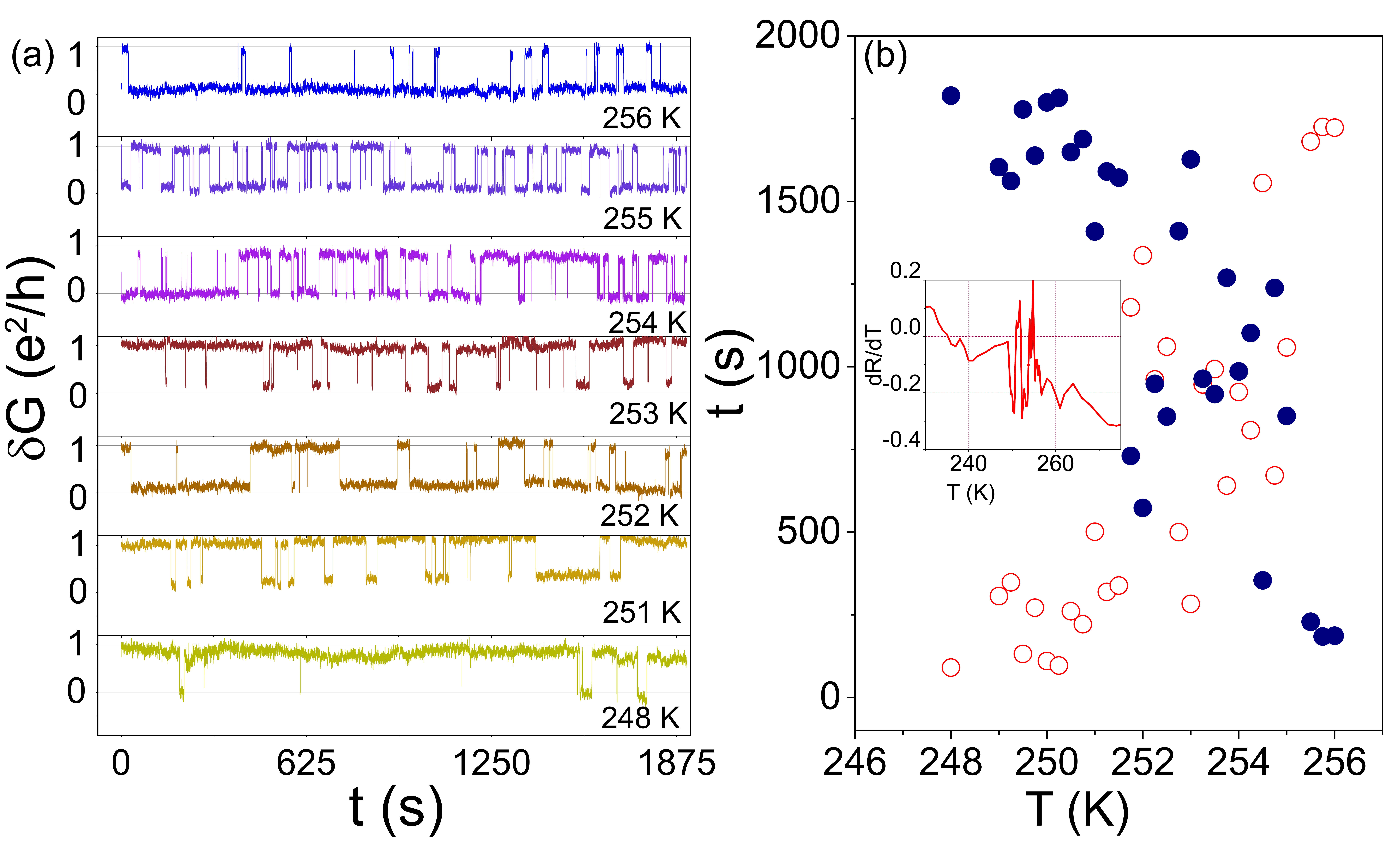}			
		\caption{(a) Plot of the time-series of conductance fluctuations measured during the  heating run. (b) Plot of time spent in the low-conductance (open red circles) and high-conductance (filled blue circles) at different $T$. The inset shows the $dR/dT$ while heating in the temperature range 230-275¬K.
			\label{Fig.4}}   
	\end{center}
\end{figure}

	From Fig.~\ref{Fig.1} we see that while heating, the system goes from a metallic phase to an insulating state at around 250K.  To study this evolution in detail, we look at the resistance fluctuations at different intermediate temperatures. Fig.~\ref{Fig.3} is a plot of the time series of resistance fluctuations measured during the heating run; the data again shows the presence of RTN. We notice that the system remains predominantly in a low-resistance state at lower temperatures, occasionally flipping to the high-resistance state.  With increasing $T$, the system spends more and more time in the higher resistance phase -- for temperatures $T>256$~K, the system remains almost exclusively higher resistance phase. From the plot of conductance fluctuations shown in Fig.~\ref{Fig.4}(a) (calculated using the relation $\delta G=\delta R/R^2$), we observe that magnitude of the conductance jump matches the quantum of conductance, $e^2/h$. In fig.~\ref{Fig.4}(b), we plot the time spent in the low-conductance (open red circles) and high-conductance (filled blue circles) at different $T$. {Inset of Fig.~\ref{Fig.4}(b) is the plot of $dR/dT$ in the temperature range 230-275~K, which shows that the system is going from a metallic to an insulating state. The system fluctuates between the two phases in the temperature range between 250~K and 260K and settles down in the insulating phase.} The PSD in the heating run was analyzed using  Eqn.~\ref{eqn.2} yielding activation energy of 496~meV. Various experiments such as ARPES, STM  studies show that this activation energy is similar to the Mott gap~\cite{lutsyk2018electronic,wang2020band}. This indicates that there is a transition from the strained metallic phase to a possible insulating Mott phase~\cite{wang2020band}. 
	
\begin{figure}[t]
		\begin{center}
			\includegraphics[width=1.0\columnwidth]{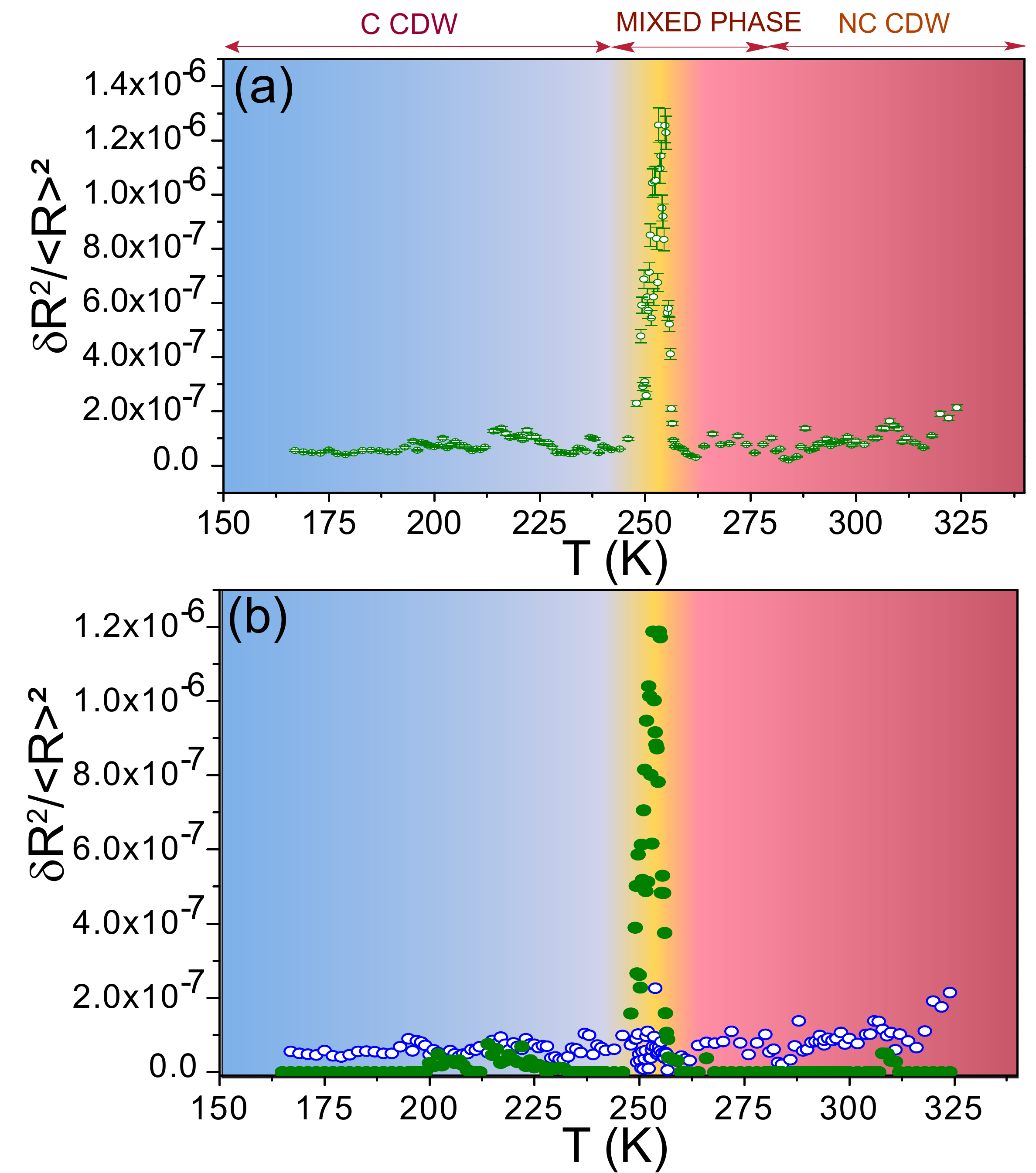}			
			\caption {(a) Relative variance of resistance fluctuations measured during the heating run.  (b) A plot of the contributions of the $1/f$ (filled green circles) and Lorentzian components (open blue circles) to the resistance fluctuations measured during the heating run.  \label{Fig.5}}   
		\end{center}
	\end{figure}

	The relative variance in the resistance fluctuations (which we refer to as `noise') was calculated by integrating $S_R(f)$ over the measured frequency bandwidth:
	\begin{equation}
		\frac{\left<\delta R^2\right>}{\left<R^2\right>}= \frac{1}{R^2}\int S_R (f)df.
		\label{eqn.3}
	\end{equation}
	{As shown in our previous experiments\cite{sarkar2019probing}, for a TMDC without CDW and high defect levels like \ch{MoS2}, the noise level at 80~K is around $10^{-5}$. For a metallic TMDC with both defect levels and CDW like \ch{NbSe2}, the noise level is $10^{-7}$ at 80~K. The noise level depends on the details of the position of the defect levels present in the system.}	The relative variance of the resistance fluctuation measured during the heating run is shown in Fig.~\ref{Fig.5}. In the region where we observe RTN, there is an increase in the relative variance by more than an order of magnitude. As shown in Fig.~\ref{Fig.5}(b), the additional noise arises from the Lorentzian component of the spectrum. Keep in mind that the peak in the relative variance (and the RTN in the conductance fluctuations) appears before the transition from C-CDW to the NC-CDW phase.{Note that we observed the C-CDW to NC-CDW transition at 286 ~K. The mixed-phase occurs at a temperature range of 248~K to 260~K. If this mixed-phase appears due to the NC-CDW transition, we would have observed the transition from C-CDW to NC-CDW immediately after the disappearance of the mixed-phase. The fact that we observe the transition to NC-CDW at 286~K (which is 26~K higher after the disappearance of the mixed-phase) leads us to conclude that this mixed phase is distinct from the mixed-phase in the NC-CDW phase observed by Yoshida et al. \cite{yoshida2015memristive}.}
	
	\begin{figure}[t]
		\begin{center}
			\includegraphics[width=1.0\columnwidth]{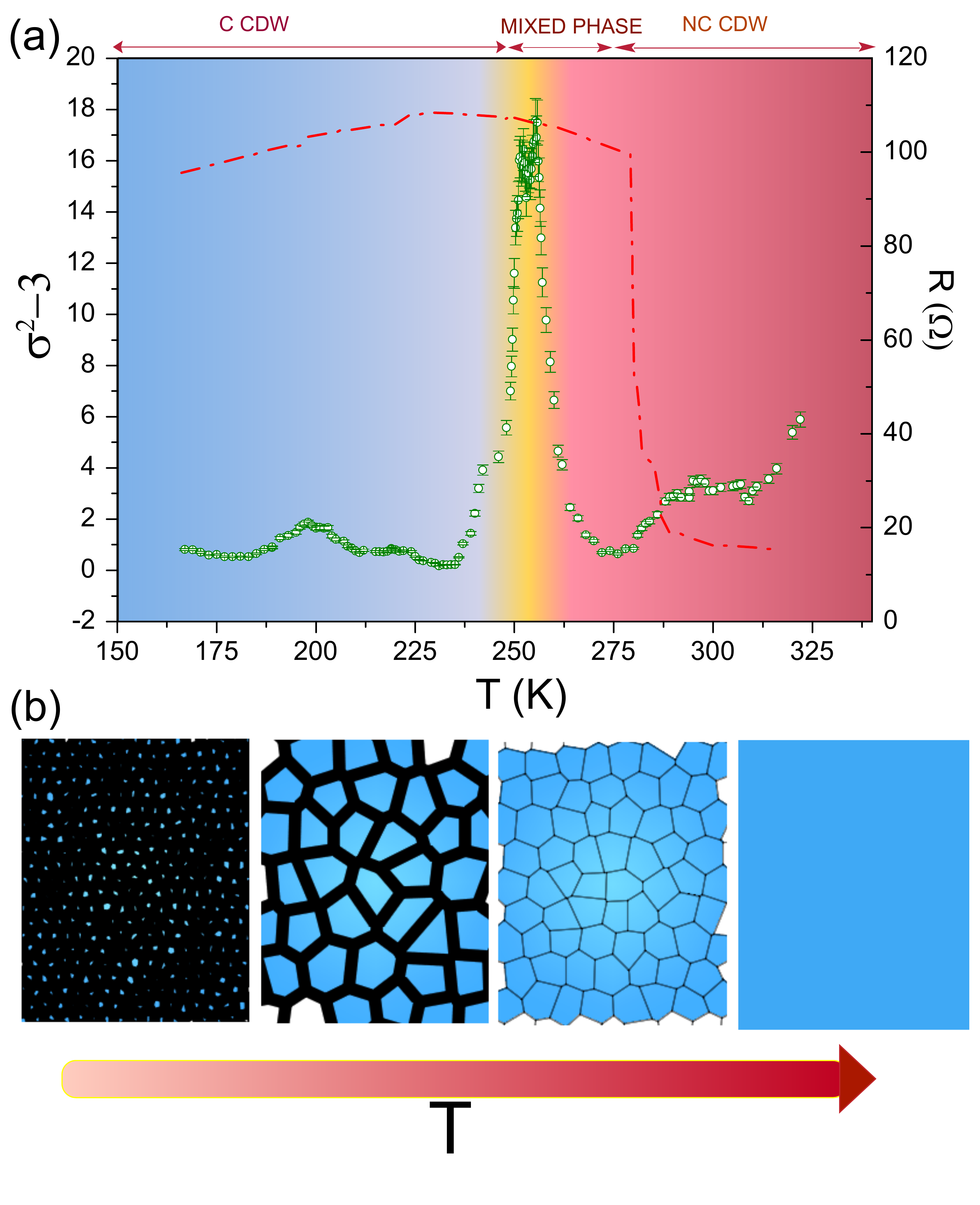}
			\caption{(a) Plot of excess second spectrum as a function of temperature heating (green open circles)in the left axis and resistance on the right axis. (b) Schematic showing the evolution of the insulating phase (light blue) from the metallic phase (black) with increasing temperature.  \label{Fig.6}}   
		\end{center}
	\end{figure}

	Combining all these observations, we present a possible phase space picture of strained 1T-\ch{TaS2}.  The resistivity data in Fig.~\ref{Fig.1} establishes that at low temperatures, the system is in a metallic phase due to the strain-induced collapse of the gap. With increasing $T$, the system goes from a high-conductance phase to a low-conductance phase with a well-defined electronically mixed-phase over an extended intermediate temperature range. To understand the origin of the mixed electronic phase, recall that the low-$T$ electronic state in 1T-\ch{TaS2} is dictated by the inter-layer interactions, which in turn, is strongly affected by lateral strain. For 1T-\ch{TaS2}, the electron at the central Ta atom of the star of David would be delocalized and expected to have a  metallic ground state. However, the dimerization of successive layers opens up a gap that drives the system into an insulating state at temperatures lower than the NC-CDW -- C-CDW transition temperature\cite{lee2019origin,ritschel2018stacking}. The nature of this low-temperature insulating phase is debated in the community, with several recent studies suggesting either a Mott insulator\cite{perfetti2005unexpected,kim1994observation}  or a simple band insulator\cite{wang2020band}. Depending on the physical model used, the insulating gap can be interpreted as either a band-gap of the band-insulator or the gap between the lower- and the upper-Hubbard bands of the Mott insulator. Proximate to this insulating phase is a putative metallic state separated by a small energy barrier\cite{lee2019origin}. 
	
	Applying a lateral strain changes the layers' relative alignment, reducing the orbital overlap between these central Ta atoms in successive layers. This destabilizes the insulating phase, and consequently, the energetically proximate metallic phase becomes the new ground state. We note that previous studies have established that TMDC flakes suspended on Au contact pads experience a strain that can be as large as 1\% at low temperatures~\cite{kundu2017quantum,PhysRevB.82.155432}. This significant strain results in the system having a metallic ground phase at low--$T$. With increasing $T$, the strain on the 1T-\ch{TaS2} decreases, leading to a change in the inter-layer orientation -- this drives the system from the metallic back into the insulating phase. Our observations match very well with the recent observation of the Mott insulator evolving from a low-temperature phase in this system ~\cite{wang2020band}.
	
	We now provide a phenomenological explanation of the occurrence of quantized conductance jumps.  As shown in the preceding paragraphs, with increasing temperature, for $T> 248$~K, percolating networks of insulating phases appear in the matrix of a metallic background (see Fig.~\ref{Fig.6}(b)). This is a dynamic situation, with thermal energy causing these mixed-phase regions to switch randomly between metallic and insulating patches. With increasing $T$, the proportion of insulating phases increase until, at around 260~K, the entire system is in a homogeneous insulating phase. It is entirely feasible that the percolating networks at higher temperatures will be primarily insulating regions separated by narrow, quasi-1D metallic domain walls~\cite{daptary2019effect,minhas2017temperature}. Thermal fluctuations shift these domain walls, leading to the making and breaking of the nodes of the percolation network -- this would cause the conductance of the mesoscopic scale devices to fluctuate by $e^2/h$. {Note that the percolation picture used here is applicable for the strained 1T-\ch{TaS2} system and may not necessarily be a general feature of 1T-\ch{TaS2}}. Further experimental and theoretical studies are necessary to validate this picture.  
	
	If there indeed is a mixed phase, as discussed in the previous paragraph, its presence will lead to long-range correlations in the system. If these couple with a degree of freedom affects the system's resistance, such correlations make themselves felt in the higher-order statistics of resistance fluctuations. According to the central limit theorem, the fluctuation statistics will be Gaussian if the fluctuators in the system are uncorrelated~\cite{reif2009fundamentals}. To probe for the presence of correlated fluctuations in the system, we calculated the `second spectrum' of resistance fluctuations~\cite{PhysRevLett.76.3049,PhysRevB.31.2254}. The second spectrum, which is extremely sensitive to the presence of the non-Gaussian component in the resistance fluctuation, is defined as the Fourier transform of the four-point correlation function of the resistance fluctuation calculated over a frequency octave $\left(f_l,f_h\right)$: 
	\begin{equation}
		S_R^{f_1}(f_2) ={\int_{0}^{\infty}<\delta R^2 (t)><\delta R^2\left( t +\tau \right) >\cos\left(2\pi f_2 \tau \right)d\tau}
		\label{eqn.4}
	\end{equation}                             
	where $f_1$ is the center frequency of the chosen octave and $f_2$ is the spectral frequency~\cite{PhysRevB.94.085104, PhysRevB.90.115153}. $S_R^{f_1}(f_2)$ physically represents the fluctuation in the PSD with time. We have chosen the octave where the sample noise is significantly higher than the background to avoid corruption of the signal by the background noise. A convenient way of representing is in its normalized form given by
	\begin{equation}
		\sigma^{\left(2\right)} = {\int_{0}^{{f_l}-{f_h}}S^{f_1}_R\left(f_2\right)df_2} \Big/{\left[\int_{f_l}^{f_h}S_R\left(f\right)df\right]}^2
		\label{eqn.5}
	\end{equation} 
	For Gaussian fluctuations, $\sigma^{(2)} = 3$; any deviation from this indicates the presence of non-Gaussian fluctuations in the system. Fig.~\ref{Fig.6}  shows the plot of excess second spectrum $\sigma^{(2)}-3$ with $T$. We can see that over the $T$ range where the RTN is present, there is an increase in the value of the second spectrum over the Gaussian value. This again indicates a significant increase in temporal correlations in the electronically mixed-phase region of the phase space. Further studies are needed to elucidate the nature of these correlations. 
	
	{We now turn to a discussion of the phase transitions occurring in the strained 1T-\ch{TaS2}. When the system is cooled down from room temperature, the NC-CDW to C-CDW transition occurs around 200~K. The C-CDW phase is insulating. On reducing the temperature further, we find the appearance of a metallic phase. We conjecture that this is because of strain in the 1T-\ch{TaS2} flake. The strain in this particular geometry arises due to the unequal thermal contraction of the 1T-\ch{TaS2} flakes and the Au probes on which they are tethered. This effect enhances with decreasing $T$; in other words, the strain increases with decreasing $T$. Recall also that for the insulating phase to be the energetically favorable phase at low temperatures, the interlayer stacking must lead to dimerization of the adjacent layers. It has been predicted that strain will reduce the interlayer interactions. This will, in turn, weaken the dimerization effect, ultimately leading to a putative metallic phase.} 
	
	{Similarly, while heating, the strain on 1T-\ch{TaS2} reduces, restoring the interlayer interactions to their pristine value. This increase in interlayer interaction will cause the system to go from metallic to the insulating phase. This is precisely the transition we observe during the heating run for 250$<T<$260~K.}
	
	\section{Conclusion}
	To conclude, we have studied the effect of tensile strain on a few-layer of 1T-\ch{TaS2}. In contrast to the unstrained system, the flakes under tensile strain transform to a metallic phase at low temperatures. Upon heating, there is a narrow range of $T$ over which the insulating phase is recovered. Preceding this, there is a range of temperatures over which the system is electronically phase separated -- as evidenced by quantized jumps in the conductance between two well-defined levels. We speculate that the quantized conductance jumps signify the movement of the metallic domain walls separating the insulating regions. Our work poses a new question -- namely, the nature of the high-temperature insulating phase. Further experimental and theoretical studies are needed to address this issue.
	
	AB acknowledges funding from SERB (HRR/2015/000017) and DST (DST/SJF/PSA-01/2016-17).  The authors acknowledge facilities in CeNSE, IISc.

	%
	

\begin{thebibliography}{38}%
		\makeatletter
		\providecommand \@ifxundefined [1]{%
			\@ifx{#1\undefined}
		}%
		\providecommand \@ifnum [1]{%
			\ifnum #1\expandafter \@firstoftwo
			\else \expandafter \@secondoftwo
			\fi
		}%
		\providecommand \@ifx [1]{%
			\ifx #1\expandafter \@firstoftwo
			\else \expandafter \@secondoftwo
			\fi
		}%
		\providecommand \natexlab [1]{#1}%
		\providecommand \enquote  [1]{``#1''}%
		\providecommand \bibnamefont  [1]{#1}%
		\providecommand \bibfnamefont [1]{#1}%
		\providecommand \citenamefont [1]{#1}%
		\providecommand \href@noop [0]{\@secondoftwo}%
		\providecommand \href [0]{\begingroup \@sanitize@url \@href}%
		\providecommand \@href[1]{\@@startlink{#1}\@@href}%
		\providecommand \@@href[1]{\endgroup#1\@@endlink}%
		\providecommand \@sanitize@url [0]{\catcode `\\12\catcode `\$12\catcode
			`\&12\catcode `\#12\catcode `\^12\catcode `\_12\catcode `\%12\relax}%
		\providecommand \@@startlink[1]{}%
		\providecommand \@@endlink[0]{}%
		\providecommand \url  [0]{\begingroup\@sanitize@url \@url }%
		\providecommand \@url [1]{\endgroup\@href {#1}{\urlprefix }}%
		\providecommand \urlprefix  [0]{URL }%
		\providecommand \Eprint [0]{\href }%
		\providecommand \doibase [0]{https://doi.org/}%
		\providecommand \selectlanguage [0]{\@gobble}%
		\providecommand \bibinfo  [0]{\@secondoftwo}%
		\providecommand \bibfield  [0]{\@secondoftwo}%
		\providecommand \translation [1]{[#1]}%
		\providecommand \BibitemOpen [0]{}%
		\providecommand \bibitemStop [0]{}%
		\providecommand \bibitemNoStop [0]{.\EOS\space}%
		\providecommand \EOS [0]{\spacefactor3000\relax}%
		\providecommand \BibitemShut  [1]{\csname bibitem#1\endcsname}%
		\let\auto@bib@innerbib\@empty
		\bibitem [{\citenamefont {Thomson}\ \emph {et~al.}(1994)\citenamefont
			{Thomson}, \citenamefont {Burk}, \citenamefont {Zettl},\ and\ \citenamefont
			{Clarke}}]{thomson1994scanning}%
		\BibitemOpen
		\bibfield  {author} {\bibinfo {author} {\bibfnamefont {R.}~\bibnamefont
				{Thomson}}, \bibinfo {author} {\bibfnamefont {B.}~\bibnamefont {Burk}},
			\bibinfo {author} {\bibfnamefont {A.}~\bibnamefont {Zettl}},\ and\ \bibinfo
			{author} {\bibfnamefont {J.}~\bibnamefont {Clarke}},\ }\bibfield  {title}
		{\bibinfo {title} {Scanning tunneling microscopy of the charge-density-wave
				structure in 1t-tas 2},\ }\href@noop {} {\bibfield  {journal} {\bibinfo
				{journal} {Physical Review B}\ }\textbf {\bibinfo {volume} {49}},\ \bibinfo
			{pages} {16899} (\bibinfo {year} {1994})}\BibitemShut {NoStop}%
		\bibitem [{\citenamefont {Wilson}\ \emph {et~al.}(1975)\citenamefont {Wilson},
			\citenamefont {Di~Salvo},\ and\ \citenamefont {Mahajan}}]{wilson1975charge}%
		\BibitemOpen
		\bibfield  {author} {\bibinfo {author} {\bibfnamefont {J.~A.}\ \bibnamefont
				{Wilson}}, \bibinfo {author} {\bibfnamefont {F.}~\bibnamefont {Di~Salvo}},\
			and\ \bibinfo {author} {\bibfnamefont {S.}~\bibnamefont {Mahajan}},\
		}\bibfield  {title} {\bibinfo {title} {Charge-density waves and superlattices
				in the metallic layered transition metal dichalcogenides},\ }\href@noop {}
		{\bibfield  {journal} {\bibinfo  {journal} {Advances in Physics}\ }\textbf
			{\bibinfo {volume} {24}},\ \bibinfo {pages} {117} (\bibinfo {year}
			{1975})}\BibitemShut {NoStop}%
		\bibitem [{\citenamefont {Rossnagel}(2011)}]{rossnagel2011origin}%
		\BibitemOpen
		\bibfield  {author} {\bibinfo {author} {\bibfnamefont {K.}~\bibnamefont
				{Rossnagel}},\ }\bibfield  {title} {\bibinfo {title} {On the origin of
				charge-density waves in select layered transition-metal dichalcogenides},\
		}\href@noop {} {\bibfield  {journal} {\bibinfo  {journal} {Journal of
					Physics: Condensed Matter}\ }\textbf {\bibinfo {volume} {23}},\ \bibinfo
			{pages} {213001} (\bibinfo {year} {2011})}\BibitemShut {NoStop}%
		\bibitem [{\citenamefont {Yoshida}\ \emph {et~al.}(2014)\citenamefont
			{Yoshida}, \citenamefont {Zhang}, \citenamefont {Ye}, \citenamefont {Suzuki},
			\citenamefont {Imai}, \citenamefont {Kimura}, \citenamefont {Fujiwara},\ and\
			\citenamefont {Iwasa}}]{yoshida2014controlling}%
		\BibitemOpen
		\bibfield  {author} {\bibinfo {author} {\bibfnamefont {M.}~\bibnamefont
				{Yoshida}}, \bibinfo {author} {\bibfnamefont {Y.}~\bibnamefont {Zhang}},
			\bibinfo {author} {\bibfnamefont {J.}~\bibnamefont {Ye}}, \bibinfo {author}
			{\bibfnamefont {R.}~\bibnamefont {Suzuki}}, \bibinfo {author} {\bibfnamefont
				{Y.}~\bibnamefont {Imai}}, \bibinfo {author} {\bibfnamefont {S.}~\bibnamefont
				{Kimura}}, \bibinfo {author} {\bibfnamefont {A.}~\bibnamefont {Fujiwara}},\
			and\ \bibinfo {author} {\bibfnamefont {Y.}~\bibnamefont {Iwasa}},\ }\bibfield
		{title} {\bibinfo {title} {Controlling charge-density-wave states in
				nano-thick crystals of 1t-tas 2},\ }\href@noop {} {\bibfield  {journal}
			{\bibinfo  {journal} {Scientific reports}\ }\textbf {\bibinfo {volume} {4}},\
			\bibinfo {pages} {1} (\bibinfo {year} {2014})}\BibitemShut {NoStop}%
		\bibitem [{\citenamefont {Yoshida}\ \emph {et~al.}(2015)\citenamefont
			{Yoshida}, \citenamefont {Suzuki}, \citenamefont {Zhang}, \citenamefont
			{Nakano},\ and\ \citenamefont {Iwasa}}]{yoshida2015memristive}%
		\BibitemOpen
		\bibfield  {author} {\bibinfo {author} {\bibfnamefont {M.}~\bibnamefont
				{Yoshida}}, \bibinfo {author} {\bibfnamefont {R.}~\bibnamefont {Suzuki}},
			\bibinfo {author} {\bibfnamefont {Y.}~\bibnamefont {Zhang}}, \bibinfo
			{author} {\bibfnamefont {M.}~\bibnamefont {Nakano}},\ and\ \bibinfo {author}
			{\bibfnamefont {Y.}~\bibnamefont {Iwasa}},\ }\bibfield  {title} {\bibinfo
			{title} {Memristive phase switching in two-dimensional 1t-tas2 crystals},\
		}\href@noop {} {\bibfield  {journal} {\bibinfo  {journal} {Science advances}\
			}\textbf {\bibinfo {volume} {1}},\ \bibinfo {pages} {e1500606} (\bibinfo
			{year} {2015})}\BibitemShut {NoStop}%
		\bibitem [{\citenamefont {Fazekas}\ and\ \citenamefont
			{Tosatti}(1979)}]{fazekas1979electrical}%
		\BibitemOpen
		\bibfield  {author} {\bibinfo {author} {\bibfnamefont {P.}~\bibnamefont
				{Fazekas}}\ and\ \bibinfo {author} {\bibfnamefont {E.}~\bibnamefont
				{Tosatti}},\ }\bibfield  {title} {\bibinfo {title} {Electrical, structural
				and magnetic properties of pure and doped 1t-tas2},\ }\href@noop {}
		{\bibfield  {journal} {\bibinfo  {journal} {Philosophical Magazine B}\
			}\textbf {\bibinfo {volume} {39}},\ \bibinfo {pages} {229} (\bibinfo {year}
			{1979})}\BibitemShut {NoStop}%
		\bibitem [{\citenamefont {Perfetti}\ \emph {et~al.}(2005)\citenamefont
			{Perfetti}, \citenamefont {Gloor}, \citenamefont {Mila}, \citenamefont
			{Berger},\ and\ \citenamefont {Grioni}}]{perfetti2005unexpected}%
		\BibitemOpen
		\bibfield  {author} {\bibinfo {author} {\bibfnamefont {L.}~\bibnamefont
				{Perfetti}}, \bibinfo {author} {\bibfnamefont {T.}~\bibnamefont {Gloor}},
			\bibinfo {author} {\bibfnamefont {F.}~\bibnamefont {Mila}}, \bibinfo {author}
			{\bibfnamefont {H.}~\bibnamefont {Berger}},\ and\ \bibinfo {author}
			{\bibfnamefont {M.}~\bibnamefont {Grioni}},\ }\bibfield  {title} {\bibinfo
			{title} {Unexpected periodicity in the quasi-two-dimensional mott insulator 1
				t- tas 2 revealed by angle-resolved photoemission},\ }\href@noop {}
		{\bibfield  {journal} {\bibinfo  {journal} {Physical Review B}\ }\textbf
			{\bibinfo {volume} {71}},\ \bibinfo {pages} {153101} (\bibinfo {year}
			{2005})}\BibitemShut {NoStop}%
		\bibitem [{\citenamefont {Kim}\ \emph {et~al.}(1994)\citenamefont {Kim},
			\citenamefont {Yamaguchi}, \citenamefont {Hasegawa},\ and\ \citenamefont
			{Kitazawa}}]{kim1994observation}%
		\BibitemOpen
		\bibfield  {author} {\bibinfo {author} {\bibfnamefont {J.-J.}\ \bibnamefont
				{Kim}}, \bibinfo {author} {\bibfnamefont {W.}~\bibnamefont {Yamaguchi}},
			\bibinfo {author} {\bibfnamefont {T.}~\bibnamefont {Hasegawa}},\ and\
			\bibinfo {author} {\bibfnamefont {K.}~\bibnamefont {Kitazawa}},\ }\bibfield
		{title} {\bibinfo {title} {Observation of mott localization gap using low
				temperature scanning tunneling spectroscopy in commensurate 1 t- t a sa 2},\
		}\href@noop {} {\bibfield  {journal} {\bibinfo  {journal} {Physical review
					letters}\ }\textbf {\bibinfo {volume} {73}},\ \bibinfo {pages} {2103}
			(\bibinfo {year} {1994})}\BibitemShut {NoStop}%
		\bibitem [{\citenamefont {Sato}\ \emph {et~al.}(2014)\citenamefont {Sato},
			\citenamefont {Arita}, \citenamefont {Utsumi}, \citenamefont {Mukaegawa},
			\citenamefont {Sasaki}, \citenamefont {Ohnishi}, \citenamefont {Kitaura},
			\citenamefont {Namatame},\ and\ \citenamefont
			{Taniguchi}}]{sato2014conduction}%
		\BibitemOpen
		\bibfield  {author} {\bibinfo {author} {\bibfnamefont {H.}~\bibnamefont
				{Sato}}, \bibinfo {author} {\bibfnamefont {M.}~\bibnamefont {Arita}},
			\bibinfo {author} {\bibfnamefont {Y.}~\bibnamefont {Utsumi}}, \bibinfo
			{author} {\bibfnamefont {Y.}~\bibnamefont {Mukaegawa}}, \bibinfo {author}
			{\bibfnamefont {M.}~\bibnamefont {Sasaki}}, \bibinfo {author} {\bibfnamefont
				{A.}~\bibnamefont {Ohnishi}}, \bibinfo {author} {\bibfnamefont
				{M.}~\bibnamefont {Kitaura}}, \bibinfo {author} {\bibfnamefont
				{H.}~\bibnamefont {Namatame}},\ and\ \bibinfo {author} {\bibfnamefont
				{M.}~\bibnamefont {Taniguchi}},\ }\bibfield  {title} {\bibinfo {title}
			{Conduction-band electronic structure of 1 t-tas 2 revealed by angle-resolved
				inverse-photoemission spectroscopy},\ }\href@noop {} {\bibfield  {journal}
			{\bibinfo  {journal} {Physical Review B}\ }\textbf {\bibinfo {volume} {89}},\
			\bibinfo {pages} {155137} (\bibinfo {year} {2014})}\BibitemShut {NoStop}%
		\bibitem [{\citenamefont {Law}\ and\ \citenamefont {Lee}(2017)}]{law20171t}%
		\BibitemOpen
		\bibfield  {author} {\bibinfo {author} {\bibfnamefont {K.~T.}\ \bibnamefont
				{Law}}\ and\ \bibinfo {author} {\bibfnamefont {P.~A.}\ \bibnamefont {Lee}},\
		}\bibfield  {title} {\bibinfo {title} {1t-tas2 as a quantum spin liquid},\
		}\href@noop {} {\bibfield  {journal} {\bibinfo  {journal} {Proceedings of the
					National Academy of Sciences}\ }\textbf {\bibinfo {volume} {114}},\ \bibinfo
			{pages} {6996} (\bibinfo {year} {2017})}\BibitemShut {NoStop}%
		\bibitem [{\citenamefont {Lee}\ \emph {et~al.}(2019)\citenamefont {Lee},
			\citenamefont {Goh},\ and\ \citenamefont {Cho}}]{lee2019origin}%
		\BibitemOpen
		\bibfield  {author} {\bibinfo {author} {\bibfnamefont {S.-H.}\ \bibnamefont
				{Lee}}, \bibinfo {author} {\bibfnamefont {J.~S.}\ \bibnamefont {Goh}},\ and\
			\bibinfo {author} {\bibfnamefont {D.}~\bibnamefont {Cho}},\ }\bibfield
		{title} {\bibinfo {title} {Origin of the insulating phase and first-order
				metal-insulator transition in 1 t- tas 2},\ }\href@noop {} {\bibfield
			{journal} {\bibinfo  {journal} {Physical review letters}\ }\textbf {\bibinfo
				{volume} {122}},\ \bibinfo {pages} {106404} (\bibinfo {year}
			{2019})}\BibitemShut {NoStop}%
		\bibitem [{\citenamefont {Ritschel}\ \emph {et~al.}(2018)\citenamefont
			{Ritschel}, \citenamefont {Berger},\ and\ \citenamefont
			{Geck}}]{ritschel2018stacking}%
		\BibitemOpen
		\bibfield  {author} {\bibinfo {author} {\bibfnamefont {T.}~\bibnamefont
				{Ritschel}}, \bibinfo {author} {\bibfnamefont {H.}~\bibnamefont {Berger}},\
			and\ \bibinfo {author} {\bibfnamefont {J.}~\bibnamefont {Geck}},\ }\bibfield
		{title} {\bibinfo {title} {Stacking-driven gap formation in layered 1t-tas
				2},\ }\href@noop {} {\bibfield  {journal} {\bibinfo  {journal} {Physical
					Review B}\ }\textbf {\bibinfo {volume} {98}},\ \bibinfo {pages} {195134}
			(\bibinfo {year} {2018})}\BibitemShut {NoStop}%
		\bibitem [{\citenamefont {Ritschel}\ \emph {et~al.}(2015)\citenamefont
			{Ritschel}, \citenamefont {Trinckauf}, \citenamefont {Koepernik},
			\citenamefont {B{\"u}chner}, \citenamefont {Zimmermann}, \citenamefont
			{Berger}, \citenamefont {Joe}, \citenamefont {Abbamonte},\ and\ \citenamefont
			{Geck}}]{ritschel2015orbital}%
		\BibitemOpen
		\bibfield  {author} {\bibinfo {author} {\bibfnamefont {T.}~\bibnamefont
				{Ritschel}}, \bibinfo {author} {\bibfnamefont {J.}~\bibnamefont {Trinckauf}},
			\bibinfo {author} {\bibfnamefont {K.}~\bibnamefont {Koepernik}}, \bibinfo
			{author} {\bibfnamefont {B.}~\bibnamefont {B{\"u}chner}}, \bibinfo {author}
			{\bibfnamefont {M.~v.}\ \bibnamefont {Zimmermann}}, \bibinfo {author}
			{\bibfnamefont {H.}~\bibnamefont {Berger}}, \bibinfo {author} {\bibfnamefont
				{Y.}~\bibnamefont {Joe}}, \bibinfo {author} {\bibfnamefont {P.}~\bibnamefont
				{Abbamonte}},\ and\ \bibinfo {author} {\bibfnamefont {J.}~\bibnamefont
				{Geck}},\ }\bibfield  {title} {\bibinfo {title} {Orbital textures and charge
				density waves in transition metal dichalcogenides},\ }\href@noop {}
		{\bibfield  {journal} {\bibinfo  {journal} {Nature physics}\ }\textbf
			{\bibinfo {volume} {11}},\ \bibinfo {pages} {328} (\bibinfo {year}
			{2015})}\BibitemShut {NoStop}%
		\bibitem [{\citenamefont {Lahoud}\ \emph {et~al.}(2014)\citenamefont {Lahoud},
			\citenamefont {Meetei}, \citenamefont {Chaska}, \citenamefont {Kanigel},\
			and\ \citenamefont {Trivedi}}]{lahoud2014emergence}%
		\BibitemOpen
		\bibfield  {author} {\bibinfo {author} {\bibfnamefont {E.}~\bibnamefont
				{Lahoud}}, \bibinfo {author} {\bibfnamefont {O.~N.}\ \bibnamefont {Meetei}},
			\bibinfo {author} {\bibfnamefont {K.}~\bibnamefont {Chaska}}, \bibinfo
			{author} {\bibfnamefont {A.}~\bibnamefont {Kanigel}},\ and\ \bibinfo {author}
			{\bibfnamefont {N.}~\bibnamefont {Trivedi}},\ }\bibfield  {title} {\bibinfo
			{title} {Emergence of a novel pseudogap metallic state in a disordered 2d
				mott insulator},\ }\href@noop {} {\bibfield  {journal} {\bibinfo  {journal}
				{Physical Review Letters}\ }\textbf {\bibinfo {volume} {112}},\ \bibinfo
			{pages} {206402} (\bibinfo {year} {2014})}\BibitemShut {NoStop}%
		\bibitem [{\citenamefont {Sipos}\ \emph {et~al.}(2008)\citenamefont {Sipos},
			\citenamefont {Kusmartseva}, \citenamefont {Akrap}, \citenamefont {Berger},
			\citenamefont {Forr{\'o}},\ and\ \citenamefont
			{Tuti{\v{s}}}}]{sipos2008mott}%
		\BibitemOpen
		\bibfield  {author} {\bibinfo {author} {\bibfnamefont {B.}~\bibnamefont
				{Sipos}}, \bibinfo {author} {\bibfnamefont {A.~F.}\ \bibnamefont
				{Kusmartseva}}, \bibinfo {author} {\bibfnamefont {A.}~\bibnamefont {Akrap}},
			\bibinfo {author} {\bibfnamefont {H.}~\bibnamefont {Berger}}, \bibinfo
			{author} {\bibfnamefont {L.}~\bibnamefont {Forr{\'o}}},\ and\ \bibinfo
			{author} {\bibfnamefont {E.}~\bibnamefont {Tuti{\v{s}}}},\ }\bibfield
		{title} {\bibinfo {title} {From mott state to superconductivity in 1t-tas
				2},\ }\href@noop {} {\bibfield  {journal} {\bibinfo  {journal} {Nature
					materials}\ }\textbf {\bibinfo {volume} {7}},\ \bibinfo {pages} {960}
			(\bibinfo {year} {2008})}\BibitemShut {NoStop}%
		\bibitem [{\citenamefont {Di~Salvo}\ \emph {et~al.}(1975)\citenamefont
			{Di~Salvo}, \citenamefont {Wilson}, \citenamefont {Bagley},\ and\
			\citenamefont {Waszczak}}]{di1975effects}%
		\BibitemOpen
		\bibfield  {author} {\bibinfo {author} {\bibfnamefont {F.}~\bibnamefont
				{Di~Salvo}}, \bibinfo {author} {\bibfnamefont {J.}~\bibnamefont {Wilson}},
			\bibinfo {author} {\bibfnamefont {B.}~\bibnamefont {Bagley}},\ and\ \bibinfo
			{author} {\bibfnamefont {J.}~\bibnamefont {Waszczak}},\ }\bibfield  {title}
		{\bibinfo {title} {Effects of doping on charge-density waves in layer
				compounds},\ }\href@noop {} {\bibfield  {journal} {\bibinfo  {journal}
				{Physical Review B}\ }\textbf {\bibinfo {volume} {12}},\ \bibinfo {pages}
			{2220} (\bibinfo {year} {1975})}\BibitemShut {NoStop}%
		\bibitem [{\citenamefont {Ang}\ \emph {et~al.}(2012)\citenamefont {Ang},
			\citenamefont {Tanaka}, \citenamefont {Ieki}, \citenamefont {Nakayama},
			\citenamefont {Sato}, \citenamefont {Li}, \citenamefont {Lu}, \citenamefont
			{Sun},\ and\ \citenamefont {Takahashi}}]{ang2012real}%
		\BibitemOpen
		\bibfield  {author} {\bibinfo {author} {\bibfnamefont {R.}~\bibnamefont
				{Ang}}, \bibinfo {author} {\bibfnamefont {Y.}~\bibnamefont {Tanaka}},
			\bibinfo {author} {\bibfnamefont {E.}~\bibnamefont {Ieki}}, \bibinfo {author}
			{\bibfnamefont {K.}~\bibnamefont {Nakayama}}, \bibinfo {author}
			{\bibfnamefont {T.}~\bibnamefont {Sato}}, \bibinfo {author} {\bibfnamefont
				{L.}~\bibnamefont {Li}}, \bibinfo {author} {\bibfnamefont {W.}~\bibnamefont
				{Lu}}, \bibinfo {author} {\bibfnamefont {Y.}~\bibnamefont {Sun}},\ and\
			\bibinfo {author} {\bibfnamefont {T.}~\bibnamefont {Takahashi}},\ }\bibfield
		{title} {\bibinfo {title} {Real-space coexistence of the melted mott state
				and superconductivity in fe-substituted 1 t- tas 2},\ }\href@noop {}
		{\bibfield  {journal} {\bibinfo  {journal} {Physical review letters}\
			}\textbf {\bibinfo {volume} {109}},\ \bibinfo {pages} {176403} (\bibinfo
			{year} {2012})}\BibitemShut {NoStop}%
		\bibitem [{\citenamefont {Stojchevska}\ \emph {et~al.}(2014)\citenamefont
			{Stojchevska}, \citenamefont {Vaskivskyi}, \citenamefont {Mertelj},
			\citenamefont {Kusar}, \citenamefont {Svetin}, \citenamefont {Brazovskii},\
			and\ \citenamefont {Mihailovic}}]{stojchevska2014ultrafast}%
		\BibitemOpen
		\bibfield  {author} {\bibinfo {author} {\bibfnamefont {L.}~\bibnamefont
				{Stojchevska}}, \bibinfo {author} {\bibfnamefont {I.}~\bibnamefont
				{Vaskivskyi}}, \bibinfo {author} {\bibfnamefont {T.}~\bibnamefont {Mertelj}},
			\bibinfo {author} {\bibfnamefont {P.}~\bibnamefont {Kusar}}, \bibinfo
			{author} {\bibfnamefont {D.}~\bibnamefont {Svetin}}, \bibinfo {author}
			{\bibfnamefont {S.}~\bibnamefont {Brazovskii}},\ and\ \bibinfo {author}
			{\bibfnamefont {D.}~\bibnamefont {Mihailovic}},\ }\bibfield  {title}
		{\bibinfo {title} {Ultrafast switching to a stable hidden quantum state in an
				electronic crystal},\ }\href@noop {} {\bibfield  {journal} {\bibinfo
				{journal} {Science}\ }\textbf {\bibinfo {volume} {344}},\ \bibinfo {pages}
			{177} (\bibinfo {year} {2014})}\BibitemShut {NoStop}%
		\bibitem [{\citenamefont {Vaskivskyi}\ \emph {et~al.}(2015)\citenamefont
			{Vaskivskyi}, \citenamefont {Gospodaric}, \citenamefont {Brazovskii},
			\citenamefont {Svetin}, \citenamefont {Sutar}, \citenamefont {Goreshnik},
			\citenamefont {Mihailovic}, \citenamefont {Mertelj},\ and\ \citenamefont
			{Mihailovic}}]{vaskivskyi2015controlling}%
		\BibitemOpen
		\bibfield  {author} {\bibinfo {author} {\bibfnamefont {I.}~\bibnamefont
				{Vaskivskyi}}, \bibinfo {author} {\bibfnamefont {J.}~\bibnamefont
				{Gospodaric}}, \bibinfo {author} {\bibfnamefont {S.}~\bibnamefont
				{Brazovskii}}, \bibinfo {author} {\bibfnamefont {D.}~\bibnamefont {Svetin}},
			\bibinfo {author} {\bibfnamefont {P.}~\bibnamefont {Sutar}}, \bibinfo
			{author} {\bibfnamefont {E.}~\bibnamefont {Goreshnik}}, \bibinfo {author}
			{\bibfnamefont {I.~A.}\ \bibnamefont {Mihailovic}}, \bibinfo {author}
			{\bibfnamefont {T.}~\bibnamefont {Mertelj}},\ and\ \bibinfo {author}
			{\bibfnamefont {D.}~\bibnamefont {Mihailovic}},\ }\bibfield  {title}
		{\bibinfo {title} {Controlling the metal-to-insulator relaxation of the
				metastable hidden quantum state in 1t-tas2},\ }\href@noop {} {\bibfield
			{journal} {\bibinfo  {journal} {Science advances}\ }\textbf {\bibinfo
				{volume} {1}},\ \bibinfo {pages} {e1500168} (\bibinfo {year}
			{2015})}\BibitemShut {NoStop}%
		\bibitem [{\citenamefont {Butler}\ \emph {et~al.}(2020)\citenamefont {Butler},
			\citenamefont {Yoshida}, \citenamefont {Hanaguri},\ and\ \citenamefont
			{Iwasa}}]{butler2020mottness}%
		\BibitemOpen
		\bibfield  {author} {\bibinfo {author} {\bibfnamefont {C.}~\bibnamefont
				{Butler}}, \bibinfo {author} {\bibfnamefont {M.}~\bibnamefont {Yoshida}},
			\bibinfo {author} {\bibfnamefont {T.}~\bibnamefont {Hanaguri}},\ and\
			\bibinfo {author} {\bibfnamefont {Y.}~\bibnamefont {Iwasa}},\ }\bibfield
		{title} {\bibinfo {title} {Mottness versus unit-cell doubling as the driver
				of the insulating state in 1 t-tas 2},\ }\href@noop {} {\bibfield  {journal}
			{\bibinfo  {journal} {Nature communications}\ }\textbf {\bibinfo {volume}
				{11}},\ \bibinfo {pages} {1} (\bibinfo {year} {2020})}\BibitemShut {NoStop}%
		\bibitem [{\citenamefont {Stahl}\ \emph {et~al.}(2020)\citenamefont {Stahl},
			\citenamefont {Kusch}, \citenamefont {Heinsch}, \citenamefont {Garbarino},
			\citenamefont {Kretzschmar}, \citenamefont {Hanff}, \citenamefont
			{Rossnagel}, \citenamefont {Geck},\ and\ \citenamefont
			{Ritschel}}]{stahl2020collapse}%
		\BibitemOpen
		\bibfield  {author} {\bibinfo {author} {\bibfnamefont {Q.}~\bibnamefont
				{Stahl}}, \bibinfo {author} {\bibfnamefont {M.}~\bibnamefont {Kusch}},
			\bibinfo {author} {\bibfnamefont {F.}~\bibnamefont {Heinsch}}, \bibinfo
			{author} {\bibfnamefont {G.}~\bibnamefont {Garbarino}}, \bibinfo {author}
			{\bibfnamefont {N.}~\bibnamefont {Kretzschmar}}, \bibinfo {author}
			{\bibfnamefont {K.}~\bibnamefont {Hanff}}, \bibinfo {author} {\bibfnamefont
				{K.}~\bibnamefont {Rossnagel}}, \bibinfo {author} {\bibfnamefont
				{J.}~\bibnamefont {Geck}},\ and\ \bibinfo {author} {\bibfnamefont
				{T.}~\bibnamefont {Ritschel}},\ }\bibfield  {title} {\bibinfo {title}
			{Collapse of layer dimerization in the photo-induced hidden state of 1t-tas
				2},\ }\href@noop {} {\bibfield  {journal} {\bibinfo  {journal} {Nature
					communications}\ }\textbf {\bibinfo {volume} {11}},\ \bibinfo {pages} {1}
			(\bibinfo {year} {2020})}\BibitemShut {NoStop}%
		\bibitem [{\citenamefont {Bu}\ \emph {et~al.}(2019)\citenamefont {Bu},
			\citenamefont {Zhang}, \citenamefont {Fei}, \citenamefont {Wu}, \citenamefont
			{Zheng}, \citenamefont {Gao}, \citenamefont {Luo}, \citenamefont {Sun},\ and\
			\citenamefont {Yin}}]{bu2019possible}%
		\BibitemOpen
		\bibfield  {author} {\bibinfo {author} {\bibfnamefont {K.}~\bibnamefont
				{Bu}}, \bibinfo {author} {\bibfnamefont {W.}~\bibnamefont {Zhang}}, \bibinfo
			{author} {\bibfnamefont {Y.}~\bibnamefont {Fei}}, \bibinfo {author}
			{\bibfnamefont {Z.}~\bibnamefont {Wu}}, \bibinfo {author} {\bibfnamefont
				{Y.}~\bibnamefont {Zheng}}, \bibinfo {author} {\bibfnamefont
				{J.}~\bibnamefont {Gao}}, \bibinfo {author} {\bibfnamefont {X.}~\bibnamefont
				{Luo}}, \bibinfo {author} {\bibfnamefont {Y.-P.}\ \bibnamefont {Sun}},\ and\
			\bibinfo {author} {\bibfnamefont {Y.}~\bibnamefont {Yin}},\ }\bibfield
		{title} {\bibinfo {title} {Possible strain induced mott gap collapse in 1
				t-tas 2},\ }\href@noop {} {\bibfield  {journal} {\bibinfo  {journal}
				{Communications Physics}\ }\textbf {\bibinfo {volume} {2}},\ \bibinfo {pages}
			{1} (\bibinfo {year} {2019})}\BibitemShut {NoStop}%
		\bibitem [{\citenamefont {Ghosh}\ \emph {et~al.}(2004)\citenamefont {Ghosh},
			\citenamefont {Kar}, \citenamefont {Bid},\ and\ \citenamefont
			{Raychaudhuri}}]{ghosh2004set}%
		\BibitemOpen
		\bibfield  {author} {\bibinfo {author} {\bibfnamefont {A.}~\bibnamefont
				{Ghosh}}, \bibinfo {author} {\bibfnamefont {S.}~\bibnamefont {Kar}}, \bibinfo
			{author} {\bibfnamefont {A.}~\bibnamefont {Bid}},\ and\ \bibinfo {author}
			{\bibfnamefont {A.}~\bibnamefont {Raychaudhuri}},\ }\bibfield  {title}
		{\bibinfo {title} {A set-up for measurement of low frequency conductance
				fluctuation (noise) using digital signal processing techniques},\ }\href@noop
		{} {\bibfield  {journal} {\bibinfo  {journal} {arXiv preprint
					cond-mat/0402130}\ } (\bibinfo {year} {2004})}\BibitemShut {NoStop}%
		\bibitem [{\citenamefont {Kundu}\ \emph {et~al.}(2017)\citenamefont {Kundu},
			\citenamefont {Ray}, \citenamefont {Dolui}, \citenamefont {Bagwe},
			\citenamefont {Choudhury}, \citenamefont {Krupanidhi}, \citenamefont {Das},
			\citenamefont {Raychaudhuri},\ and\ \citenamefont {Bid}}]{kundu2017quantum}%
		\BibitemOpen
		\bibfield  {author} {\bibinfo {author} {\bibfnamefont {H.~K.}\ \bibnamefont
				{Kundu}}, \bibinfo {author} {\bibfnamefont {S.}~\bibnamefont {Ray}}, \bibinfo
			{author} {\bibfnamefont {K.}~\bibnamefont {Dolui}}, \bibinfo {author}
			{\bibfnamefont {V.}~\bibnamefont {Bagwe}}, \bibinfo {author} {\bibfnamefont
				{P.~R.}\ \bibnamefont {Choudhury}}, \bibinfo {author} {\bibfnamefont
				{S.}~\bibnamefont {Krupanidhi}}, \bibinfo {author} {\bibfnamefont
				{T.}~\bibnamefont {Das}}, \bibinfo {author} {\bibfnamefont {P.}~\bibnamefont
				{Raychaudhuri}},\ and\ \bibinfo {author} {\bibfnamefont {A.}~\bibnamefont
				{Bid}},\ }\bibfield  {title} {\bibinfo {title} {Quantum phase transition in
				few-layer nbse 2 probed through quantized conductance fluctuations},\
		}\href@noop {} {\bibfield  {journal} {\bibinfo  {journal} {Physical review
					letters}\ }\textbf {\bibinfo {volume} {119}},\ \bibinfo {pages} {226802}
			(\bibinfo {year} {2017})}\BibitemShut {NoStop}%
		\bibitem [{\citenamefont {Sengupta}\ \emph {et~al.}(2010)\citenamefont
			{Sengupta}, \citenamefont {Solanki}, \citenamefont {Singh}, \citenamefont
			{Dhara},\ and\ \citenamefont {Deshmukh}}]{PhysRevB.82.155432}%
		\BibitemOpen
		\bibfield  {author} {\bibinfo {author} {\bibfnamefont {S.}~\bibnamefont
				{Sengupta}}, \bibinfo {author} {\bibfnamefont {H.~S.}\ \bibnamefont
				{Solanki}}, \bibinfo {author} {\bibfnamefont {V.}~\bibnamefont {Singh}},
			\bibinfo {author} {\bibfnamefont {S.}~\bibnamefont {Dhara}},\ and\ \bibinfo
			{author} {\bibfnamefont {M.~M.}\ \bibnamefont {Deshmukh}},\ }\bibfield
		{title} {\bibinfo {title} {Electromechanical resonators as probes of the
				charge density wave transition at the nanoscale in ${\text{nbse}}_{2}$},\
		}\href {https://doi.org/10.1103/PhysRevB.82.155432} {\bibfield  {journal}
			{\bibinfo  {journal} {Phys. Rev. B}\ }\textbf {\bibinfo {volume} {82}},\
			\bibinfo {pages} {155432} (\bibinfo {year} {2010})}\BibitemShut {NoStop}%
		\bibitem [{\citenamefont {Tsen}\ \emph {et~al.}(2015)\citenamefont {Tsen},
			\citenamefont {Hovden}, \citenamefont {Wang}, \citenamefont {Kim},
			\citenamefont {Okamoto}, \citenamefont {Spoth}, \citenamefont {Liu},
			\citenamefont {Lu}, \citenamefont {Sun}, \citenamefont {Hone} \emph
			{et~al.}}]{tsen2015structure}%
		\BibitemOpen
		\bibfield  {author} {\bibinfo {author} {\bibfnamefont {A.~W.}\ \bibnamefont
				{Tsen}}, \bibinfo {author} {\bibfnamefont {R.}~\bibnamefont {Hovden}},
			\bibinfo {author} {\bibfnamefont {D.}~\bibnamefont {Wang}}, \bibinfo {author}
			{\bibfnamefont {Y.~D.}\ \bibnamefont {Kim}}, \bibinfo {author} {\bibfnamefont
				{J.}~\bibnamefont {Okamoto}}, \bibinfo {author} {\bibfnamefont {K.~A.}\
				\bibnamefont {Spoth}}, \bibinfo {author} {\bibfnamefont {Y.}~\bibnamefont
				{Liu}}, \bibinfo {author} {\bibfnamefont {W.}~\bibnamefont {Lu}}, \bibinfo
			{author} {\bibfnamefont {Y.}~\bibnamefont {Sun}}, \bibinfo {author}
			{\bibfnamefont {J.~C.}\ \bibnamefont {Hone}}, \emph {et~al.},\ }\bibfield
		{title} {\bibinfo {title} {Structure and control of charge density waves in
				two-dimensional 1t-tas2},\ }\href@noop {} {\bibfield  {journal} {\bibinfo
				{journal} {Proceedings of the National Academy of Sciences}\ }\textbf
			{\bibinfo {volume} {112}},\ \bibinfo {pages} {15054} (\bibinfo {year}
			{2015})}\BibitemShut {NoStop}%
		\bibitem [{\citenamefont {Bid}\ \emph {et~al.}(2003)\citenamefont {Bid},
			\citenamefont {Guha},\ and\ \citenamefont {Raychaudhuri}}]{bid2003low}%
		\BibitemOpen
		\bibfield  {author} {\bibinfo {author} {\bibfnamefont {A.}~\bibnamefont
				{Bid}}, \bibinfo {author} {\bibfnamefont {A.}~\bibnamefont {Guha}},\ and\
			\bibinfo {author} {\bibfnamefont {A.}~\bibnamefont {Raychaudhuri}},\
		}\bibfield  {title} {\bibinfo {title} {Low-frequency random telegraphic noise
				and 1/f noise in the rare-earth manganite pr 0.63 ca 0.37 mno 3 near the
				charge-ordering transition},\ }\href@noop {} {\bibfield  {journal} {\bibinfo
				{journal} {Physical Review B}\ }\textbf {\bibinfo {volume} {67}},\ \bibinfo
			{pages} {174415} (\bibinfo {year} {2003})}\BibitemShut {NoStop}%
		\bibitem [{\citenamefont {Amin}\ and\ \citenamefont
			{Bid}(2015)}]{amin2015effect}%
		\BibitemOpen
		\bibfield  {author} {\bibinfo {author} {\bibfnamefont {K.~R.}\ \bibnamefont
				{Amin}}\ and\ \bibinfo {author} {\bibfnamefont {A.}~\bibnamefont {Bid}},\
		}\bibfield  {title} {\bibinfo {title} {Effect of ambient on the resistance
				fluctuations of graphene},\ }\href@noop {} {\bibfield  {journal} {\bibinfo
				{journal} {Applied Physics Letters}\ }\textbf {\bibinfo {volume} {106}},\
			\bibinfo {pages} {183105} (\bibinfo {year} {2015})}\BibitemShut {NoStop}%
		\bibitem [{\citenamefont {Lutsyk}\ \emph {et~al.}(2018)\citenamefont {Lutsyk},
			\citenamefont {Rogala}, \citenamefont {Dabrowski}, \citenamefont {Krukowski},
			\citenamefont {Kowalczyk}, \citenamefont {Busiakiewicz}, \citenamefont
			{Kowalczyk}, \citenamefont {Lacinska}, \citenamefont {Binder}, \citenamefont
			{Olszowska} \emph {et~al.}}]{lutsyk2018electronic}%
		\BibitemOpen
		\bibfield  {author} {\bibinfo {author} {\bibfnamefont {I.}~\bibnamefont
				{Lutsyk}}, \bibinfo {author} {\bibfnamefont {M.}~\bibnamefont {Rogala}},
			\bibinfo {author} {\bibfnamefont {P.}~\bibnamefont {Dabrowski}}, \bibinfo
			{author} {\bibfnamefont {P.}~\bibnamefont {Krukowski}}, \bibinfo {author}
			{\bibfnamefont {P.}~\bibnamefont {Kowalczyk}}, \bibinfo {author}
			{\bibfnamefont {A.}~\bibnamefont {Busiakiewicz}}, \bibinfo {author}
			{\bibfnamefont {D.}~\bibnamefont {Kowalczyk}}, \bibinfo {author}
			{\bibfnamefont {E.}~\bibnamefont {Lacinska}}, \bibinfo {author}
			{\bibfnamefont {J.}~\bibnamefont {Binder}}, \bibinfo {author} {\bibfnamefont
				{N.}~\bibnamefont {Olszowska}}, \emph {et~al.},\ }\bibfield  {title}
		{\bibinfo {title} {Electronic structure of commensurate, nearly commensurate,
				and incommensurate phases of 1 t- ta s 2 by angle-resolved photoelectron
				spectroscopy, scanning tunneling spectroscopy, and density functional
				theory},\ }\href@noop {} {\bibfield  {journal} {\bibinfo  {journal} {Physical
					Review B}\ }\textbf {\bibinfo {volume} {98}},\ \bibinfo {pages} {195425}
			(\bibinfo {year} {2018})}\BibitemShut {NoStop}%
		\bibitem [{\citenamefont {Wang}\ \emph {et~al.}(2020)\citenamefont {Wang},
			\citenamefont {Yao}, \citenamefont {Xin}, \citenamefont {Han}, \citenamefont
			{Wang}, \citenamefont {Chen}, \citenamefont {Cai}, \citenamefont {Li},\ and\
			\citenamefont {Zhang}}]{wang2020band}%
		\BibitemOpen
		\bibfield  {author} {\bibinfo {author} {\bibfnamefont {Y.}~\bibnamefont
				{Wang}}, \bibinfo {author} {\bibfnamefont {W.}~\bibnamefont {Yao}}, \bibinfo
			{author} {\bibfnamefont {Z.}~\bibnamefont {Xin}}, \bibinfo {author}
			{\bibfnamefont {T.}~\bibnamefont {Han}}, \bibinfo {author} {\bibfnamefont
				{Z.}~\bibnamefont {Wang}}, \bibinfo {author} {\bibfnamefont {L.}~\bibnamefont
				{Chen}}, \bibinfo {author} {\bibfnamefont {C.}~\bibnamefont {Cai}}, \bibinfo
			{author} {\bibfnamefont {Y.}~\bibnamefont {Li}},\ and\ \bibinfo {author}
			{\bibfnamefont {Y.}~\bibnamefont {Zhang}},\ }\bibfield  {title} {\bibinfo
			{title} {Band insulator to mott insulator transition in 1 t-tas 2},\
		}\href@noop {} {\bibfield  {journal} {\bibinfo  {journal} {Nature
					communications}\ }\textbf {\bibinfo {volume} {11}},\ \bibinfo {pages} {1}
			(\bibinfo {year} {2020})}\BibitemShut {NoStop}%
		\bibitem [{\citenamefont {Sarkar}\ \emph {et~al.}(2019)\citenamefont {Sarkar},
			\citenamefont {Bid}, \citenamefont {Ganapathi},\ and\ \citenamefont
			{Mohan}}]{sarkar2019probing}%
		\BibitemOpen
		\bibfield  {author} {\bibinfo {author} {\bibfnamefont {S.}~\bibnamefont
				{Sarkar}}, \bibinfo {author} {\bibfnamefont {A.}~\bibnamefont {Bid}},
			\bibinfo {author} {\bibfnamefont {K.~L.}\ \bibnamefont {Ganapathi}},\ and\
			\bibinfo {author} {\bibfnamefont {S.}~\bibnamefont {Mohan}},\ }\bibfield
		{title} {\bibinfo {title} {Probing defect states in few-layer mos 2 by
				conductance fluctuation spectroscopy},\ }\href@noop {} {\bibfield  {journal}
			{\bibinfo  {journal} {Physical Review B}\ }\textbf {\bibinfo {volume} {99}},\
			\bibinfo {pages} {245419} (\bibinfo {year} {2019})}\BibitemShut {NoStop}%
		\bibitem [{\citenamefont {Daptary}\ \emph {et~al.}(2019)\citenamefont
			{Daptary}, \citenamefont {Kundu}, \citenamefont {Kumar}, \citenamefont
			{Dogra}, \citenamefont {Mohanta}, \citenamefont {Taraphder},\ and\
			\citenamefont {Bid}}]{daptary2019effect}%
		\BibitemOpen
		\bibfield  {author} {\bibinfo {author} {\bibfnamefont {G.~N.}\ \bibnamefont
				{Daptary}}, \bibinfo {author} {\bibfnamefont {H.~K.}\ \bibnamefont {Kundu}},
			\bibinfo {author} {\bibfnamefont {P.}~\bibnamefont {Kumar}}, \bibinfo
			{author} {\bibfnamefont {A.}~\bibnamefont {Dogra}}, \bibinfo {author}
			{\bibfnamefont {N.}~\bibnamefont {Mohanta}}, \bibinfo {author} {\bibfnamefont
				{A.}~\bibnamefont {Taraphder}},\ and\ \bibinfo {author} {\bibfnamefont
				{A.}~\bibnamefont {Bid}},\ }\bibfield  {title} {\bibinfo {title} {Effect of
				spin-orbit interaction on the vortex dynamics in laalo 3/srtio 3 interfaces
				near the superconducting transition},\ }\href@noop {} {\bibfield  {journal}
			{\bibinfo  {journal} {Physical Review B}\ }\textbf {\bibinfo {volume}
				{100}},\ \bibinfo {pages} {125117} (\bibinfo {year} {2019})}\BibitemShut
		{NoStop}%
		\bibitem [{\citenamefont {Minhas}\ \emph {et~al.}(2017)\citenamefont {Minhas},
			\citenamefont {M{\"u}ller}, \citenamefont {Heyroth}, \citenamefont
			{Blaschek},\ and\ \citenamefont {Schmidt}}]{minhas2017temperature}%
		\BibitemOpen
		\bibfield  {author} {\bibinfo {author} {\bibfnamefont {M.}~\bibnamefont
				{Minhas}}, \bibinfo {author} {\bibfnamefont {A.}~\bibnamefont {M{\"u}ller}},
			\bibinfo {author} {\bibfnamefont {F.}~\bibnamefont {Heyroth}}, \bibinfo
			{author} {\bibfnamefont {H.}~\bibnamefont {Blaschek}},\ and\ \bibinfo
			{author} {\bibfnamefont {G.}~\bibnamefont {Schmidt}},\ }\bibfield  {title}
		{\bibinfo {title} {Temperature dependent giant resistance anomaly in laalo
				3/srtio 3 nanostructures},\ }\href@noop {} {\bibfield  {journal} {\bibinfo
				{journal} {Scientific reports}\ }\textbf {\bibinfo {volume} {7}},\ \bibinfo
			{pages} {1} (\bibinfo {year} {2017})}\BibitemShut {NoStop}%
		\bibitem [{\citenamefont {Reif}(2009)}]{reif2009fundamentals}%
		\BibitemOpen
		\bibfield  {author} {\bibinfo {author} {\bibfnamefont {F.}~\bibnamefont
				{Reif}},\ }\href@noop {} {\emph {\bibinfo {title} {Fundamentals of
					statistical and thermal physics}}}\ (\bibinfo  {publisher} {Waveland Press},\
		\bibinfo {year} {2009})\BibitemShut {NoStop}%
		\bibitem [{\citenamefont {Seidler}\ \emph {et~al.}(1996)\citenamefont
			{Seidler}, \citenamefont {Solin},\ and\ \citenamefont
			{Marley}}]{PhysRevLett.76.3049}%
		\BibitemOpen
		\bibfield  {author} {\bibinfo {author} {\bibfnamefont {G.~T.}\ \bibnamefont
				{Seidler}}, \bibinfo {author} {\bibfnamefont {S.~A.}\ \bibnamefont {Solin}},\
			and\ \bibinfo {author} {\bibfnamefont {A.~C.}\ \bibnamefont {Marley}},\
		}\bibfield  {title} {\bibinfo {title} {Dynamical current redistribution and
				non-gaussian 1 $/$f noise},\ }\href
		{https://doi.org/10.1103/PhysRevLett.76.3049} {\bibfield  {journal} {\bibinfo
				{journal} {Phys. Rev. Lett.}\ }\textbf {\bibinfo {volume} {76}},\ \bibinfo
			{pages} {3049} (\bibinfo {year} {1996})}\BibitemShut {NoStop}%
		\bibitem [{\citenamefont {Restle}\ \emph {et~al.}(1985)\citenamefont {Restle},
			\citenamefont {Hamilton}, \citenamefont {Weissman},\ and\ \citenamefont
			{Love}}]{PhysRevB.31.2254}%
		\BibitemOpen
		\bibfield  {author} {\bibinfo {author} {\bibfnamefont {P.~J.}\ \bibnamefont
				{Restle}}, \bibinfo {author} {\bibfnamefont {R.~J.}\ \bibnamefont
				{Hamilton}}, \bibinfo {author} {\bibfnamefont {M.~B.}\ \bibnamefont
				{Weissman}},\ and\ \bibinfo {author} {\bibfnamefont {M.~S.}\ \bibnamefont
				{Love}},\ }\bibfield  {title} {\bibinfo {title} {Non-gaussian effects in 1/f
				noise in small silicon-on-sapphire resistors},\ }\href
		{https://doi.org/10.1103/PhysRevB.31.2254} {\bibfield  {journal} {\bibinfo
				{journal} {Phys. Rev. B}\ }\textbf {\bibinfo {volume} {31}},\ \bibinfo
			{pages} {2254} (\bibinfo {year} {1985})}\BibitemShut {NoStop}%
		\bibitem [{\citenamefont {Daptary}\ \emph {et~al.}(2016)\citenamefont
			{Daptary}, \citenamefont {Kumar}, \citenamefont {Kumar}, \citenamefont
			{Dogra}, \citenamefont {Mohanta}, \citenamefont {Taraphder},\ and\
			\citenamefont {Bid}}]{PhysRevB.94.085104}%
		\BibitemOpen
		\bibfield  {author} {\bibinfo {author} {\bibfnamefont {G.~N.}\ \bibnamefont
				{Daptary}}, \bibinfo {author} {\bibfnamefont {S.}~\bibnamefont {Kumar}},
			\bibinfo {author} {\bibfnamefont {P.}~\bibnamefont {Kumar}}, \bibinfo
			{author} {\bibfnamefont {A.}~\bibnamefont {Dogra}}, \bibinfo {author}
			{\bibfnamefont {N.}~\bibnamefont {Mohanta}}, \bibinfo {author} {\bibfnamefont
				{A.}~\bibnamefont {Taraphder}},\ and\ \bibinfo {author} {\bibfnamefont
				{A.}~\bibnamefont {Bid}},\ }\bibfield  {title} {\bibinfo {title} {Correlated
				non-gaussian phase fluctuations in
				${\mathrm{laalo}}_{3}/{\mathrm{srtio}}_{3}$ heterointerfaces},\ }\href
		{https://doi.org/10.1103/PhysRevB.94.085104} {\bibfield  {journal} {\bibinfo
				{journal} {Phys. Rev. B}\ }\textbf {\bibinfo {volume} {94}},\ \bibinfo
			{pages} {085104} (\bibinfo {year} {2016})}\BibitemShut {NoStop}%
		\bibitem [{\citenamefont {Daptary}\ \emph {et~al.}(2014)\citenamefont
			{Daptary}, \citenamefont {Sow}, \citenamefont {Kumar},\ and\ \citenamefont
			{Bid}}]{PhysRevB.90.115153}%
		\BibitemOpen
		\bibfield  {author} {\bibinfo {author} {\bibfnamefont {G.~N.}\ \bibnamefont
				{Daptary}}, \bibinfo {author} {\bibfnamefont {C.}~\bibnamefont {Sow}},
			\bibinfo {author} {\bibfnamefont {P.~S.~A.}\ \bibnamefont {Kumar}},\ and\
			\bibinfo {author} {\bibfnamefont {A.}~\bibnamefont {Bid}},\ }\bibfield
		{title} {\bibinfo {title} {Probing a spin-glass state in
				${\mathrm{srruo}}_{3}$ thin films through higher-order statistics of
				resistance fluctuations},\ }\href
		{https://doi.org/10.1103/PhysRevB.90.115153} {\bibfield  {journal} {\bibinfo
				{journal} {Phys. Rev. B}\ }\textbf {\bibinfo {volume} {90}},\ \bibinfo
			{pages} {115153} (\bibinfo {year} {2014})}\BibitemShut {NoStop}%
	\end{thebibliography}
	
\end{document}